\documentclass[10pt, letterpaper, oneside]{article}
\usepackage[utf8]{inputenc}
\usepackage[margin=1.1in]{geometry}
\usepackage{textcomp}
\usepackage{graphicx}
\usepackage{placeins}
\usepackage{amsfonts}
\usepackage{amsmath}
\usepackage{upgreek}
\usepackage{gensymb}
\usepackage[round]{natbib}
\usepackage{natbib}
\usepackage[onehalfspacing]{setspace}
\usepackage[colorlinks=true, allcolors=black, citecolor=blue]{hyperref}
\usepackage{lineno}
\usepackage[format=plain, labelfont=it, textfont=it]{caption}
\usepackage{pdflscape}
            
\DeclareUnicodeCharacter{200B}{I~AM~HERE!!!!}
\DeclareMathOperator*{\argmin}{arg\,min}

\title{
	\textbf{Improved reproducibility of diffusion kurtosis imaging using regularized non-linear optimization informed by artificial neural networks}
}

\author{
	Leevi Kerkelä$^1$*, Kiran Seunarine$^1$, Rafael Neto Henriques$^2$, \\ Jonathan D. Clayden$^1$, and Chris A. Clark$^1$
}

\date{
	\begin{flushleft}
	\scriptsize{
	    $^1$ UCL Great Ormond Street Institute of Child Health, University College London, London, UK \\
		$^2$ Champalimaud Research, Champalimaud Centre for the Unknown, Lisbon, Portugal \\
		[2ex]
		* Corresponding author: leevi.kerkela.17@ucl.ac.uk; Developmental Imaging \& Biophysics Section, UCL Great Ormond Street Institute of Child Health, 30 Guilford Street, WC1N 1EH, London, UK
	}
    \end{flushleft}
	\normalsize{\today}
}

\begin{document}

\pagenumbering{roman}

\maketitle

\newpage

\section*{Abstract}

Diffusion kurtosis imaging is an extension of diffusion tensor imaging that provides scientifically and clinically valuable information about brain tissue microstructure but suffers from poor robustness to noise, especially in voxels containing tightly packed aligned axons. We present a new algorithm for estimating diffusion and kurtosis tensors using regularized non-linear optimization and make it publicly available in an easy-to-use open-source Python software package. Our approach uses fully-connected feed-forward neural networks to predict kurtosis values in voxels where the standard non-linear least squares fit fails. The predicted values are then used in the objective function to avoid implausible kurtosis values. We show that our algorithm is more robust than standard non-linear least squares and a previously proposed regularized non-linear optimization method. The algorithm was then applied on a multi-site scan-rescan dataset acquired using a clinical scan protocol to assess the reproducibility of diffusion kurtosis parameter estimation in human white matter using the proposed algorithm. Our results show that the reproducibility of diffusion kurtosis parameters is similar to diffusion tensor parameters.

\section*{Keywords}

Diffusion kurtosis imaging; Machine learning; Non-linear optimization; Reproducibility

\section*{Abbreviations}

AD, axial diffusivity; ADC, apparent diffusion coefficient; AK, axial kurtosis; AKC, apparent kurtosis coefficient; BFGS, Broyden-Fletcher-Goldfarb-Shanno; CV, coefficient of variation; 	DKI, diffusion kurtosis imaging; dMRI, diffusion-weighted magnetic resonance imaging; DTI, diffusion tensor imaging; MD, mean diffusivity; MK, mean kurtosis; NLLS, non-linear least squares; PGSE, pulsed gradient spin echo; RD, radial diffusivity; RK, radial kurtosis

\newpage

\pagenumbering{arabic}


\section{Introduction}

Diffusion-weighted magnetic resonance imaging (dMRI) measures the displacements of water molecules on a microscopic level and thus can be sensitive to changes in brain tissue microstructure \citep{johansen2013diffusion}. Several studies have demonstrated dMRI's sensitivity to microstructural changes in the brain associated with normal development during childhood \citep{barnea2005white}, ageing \citep{grieve2007cognitive}, learning \citep{blumenfeld2011diffusion}, and neurodegenerative diseases \citep{zhang2009white}, to name a few examples. dMRI is limited by noise in signal and artefacts that can significantly reduce the accuracy and precision of microstructural parameter estimates. Moreover, in order to measure small effects, studies with a large number of participants combining data from multiple scan sites may be needed.

Over the previous two decades, scanners with strong enough magnetic fields and gradients to support 2-shell acquisition protocols have become increasingly common in both clinical and research settings \citep{van2013wu, shafto2014cambridge}. Such protocols measure the diffusion-weighted signal at two different non-zero b-values, providing more information on tissue microstructure than what conventional diffusion tensor imaging (DTI) parameters contain. Diffusion kurtosis imaging (DKI) \citep{jensen2005diffusional} is an extension of DTI that quantifies how much the diffusion propagator deviates from a multi-variate normal (i.e., Gaussian) distribution. Despite having been shown to be sensitive to clinically and scientifically relevant microstructural properties of brain tissue \citep{falangola2008age, hui2012stroke, price2017age, van2012gliomas, zhuo2012diffusion}, kurtosis tensor estimation using least squares suffers from poor robustness, especially in voxels containing tightly packed aligned axons, such as the genu and splenium of the corpus callosum, where radial diffusivity is very low. In these voxels, linear and non-linear least squares often produce implausibly low or negative values of diffusional kurtosis \citep{tabesh2011estimation, kuder2012advanced, neto2018advanced}.

Recently, \cite{henriques2021toward} showed that the robustness of DKI can be improved by using regularized least squares. They used polynomial regression to predict mean kurtosis (MK) from the estimated diffusion tensor elements and powder kurtosis (i.e., apparent kurtosis coefficient estimated from data averaged over the acquired diffusion encoding directions), and added a regularization term to the objective function with the aim of preventing the estimated parameters from taking values for which MK deviates far from the predicted value. Indeed, their method produces a significantly smaller number of voxels with incorrect negative values than standard non-linear least squares (NLLS). However, if the regularization term only depends on MK, and not also on axial kurtosis (AK) and radial kurtosis (RK), we hypothesized that it may not sufficiently penalize the parameters for taking values for which only RK is negative. This is because, unlike with the diffusion tensor, MK is independent of AK and RK \citep{hansen2016fast}.

The aim of this study was to assess the reproducibility of DKI in a clinical setting using regularized least squares. We implemented a novel algorithm for estimating diffusion and kurtosis tensors and made it publicly available in an easy-to-use open-source software package.  Our algorithm uses artificial neural networks to predict MK, AK, and RK directly from data by using a training dataset that consists of voxels where the standard NLLS fit produced only non-negative values of diffusional kurtosis. The predicted values are then used in the objective function to regularize the fit in order to avoid implausible kurtosis tensor estimates. After showing the advantages of our approach compared to the one proposed by \cite{henriques2021toward}, we investigate the reproducibility of kurtosis parameter estimates in white matter by applying the developed software on a multi-site scan-rescan dataset acquired with an imaging protocol regularly used in a clinical setting.

\section{Theory}

Using the cumulant expansion \citep{kiselev2010cumulant} until the second order in terms of the applied b-value, the diffusion-weighted magnetic resonance signal acquired with conventional pulsed gradient spin echo (PGSE) sequences can be expressed as
\begin{equation}\label{eq:cumulant_expansion}
\begin{split}
S(\mathbf{n}, b) = S_0 \exp \Bigg( & - b \sum_{i=1}^3 \sum_{j=1}^3 n_i n_j D_{ij} \\ & + \frac{1}{6} b^2 \left(\text{MD}\right)^2 \sum_{i=1}^3 \sum_{j=1}^3 \sum_{k=1}^3 \sum_{l=1}^3 n_i n_j n_k n_l W_{ijkl} \Bigg) ,
\end{split}
\end{equation}
where $\mathbf{n}$ is a unit vector aligned with the diffusion encoding direction, $b$ is the b-value, $S_0$ is the signal magnitude without diffusion-weighting, $D_{ij}$ is an element of the diffusion tensor $\mathbf{D}$ \citep{basser1994mr}, $\text{MD} = \text{Tr}\left(\mathbf{D}\right) / 3$ is mean diffusivity, and $W_{ijkl}$ is an element of the kurtosis tensor $\mathbf{W}$ \citep{jensen2005diffusional}. $\mathbf{D}$ is a symmetric rank-3 tensor with shape $3 \times 3$ and 6 unique elements, and $\mathbf{W}$ is a symmetric rank-4  tensor with shape $3 \times 3 \times 3 \times 3$ and 15 unique elements.

Several scalar metrics quantifying relevant properties of $\mathbf{W}$ have been introduced \citep{jensen2010mri}. Apparent kurtosis coefficient along $\mathbf{n}$ is
\begin{equation}\label{eq:akc}
\text{AKC}(\mathbf{n}) \equiv \frac{\left(\text{MD}\right)^2}{\left(\text{ADC}(\mathbf{n})\right)^2} n_i n_j n_k n_l W_{ijkl} ,
\end{equation}
where $\text{ADC}(\mathbf{n}) = n_i n_j D_{ij}$ is the apparent diffusion coefficient along $\mathbf{n}$. Mean kurtosis is
\begin{equation}\label{eq:mk}
\text{MK} \equiv \frac{1}{4\pi}\int d{\Omega}_\mathbf{n} \text{AKC}(\mathbf{n}) ,
\end{equation}
where $d{\Omega}_\mathbf{n}$ denotes a solid angle element. Axial kurtosis is
\begin{equation}\label{eq:ak}
\text{AK} \equiv \text{AKC}\left(\mathbf{e}_1\right) ,
\end{equation}
where $\mathbf{e}_1$ is the eigenvector corresponding to the largest eigenvalue of $\mathbf{D}$ (i.e., principal diffusion direction). Radial kurtosis is
\begin{equation}\label{eq:rk}
\text{RK} \equiv \frac{1}{4\pi}\int d{\Omega}_\mathbf{n} \text{AKC}(\mathbf{n})\delta(\mathbf{n} \cdot \mathbf{e}_1) ,
\end{equation}
where $\delta$ is the Dirac delta function.

\section{Methods}

\subsection{Parameter estimation}

The right-hand side of Equation \ref{eq:cumulant_expansion} was rewritten as $\exp \left( \mathbf{X} \boldsymbol{\beta} \right)$, where $\mathbf{X}$ is the design matrix and $\boldsymbol{\beta}$ is a column vector containing the model parameters $\ln \left (S_0 \right)$, 6 unique elements of $\mathbf{D}$, and 15 unique elements of $\mathbf{W}$ multiplied by $\left( \text{MD} \right)^2$ \citep{tabesh2011estimation}. The model parameters were estimated using an algorithm that consists of the the steps explained below. All computation was performed using Python software package that is publicly available at \url{https://github.com/kerkelae/dkmri}. Python code was accelerated by using JAX \citep{jax2018github} and Numba \citep{lam2015numba}.

\subsubsection{Standard non-linear least squares}\label{step_1}

The following optimization problem was solved using the BFGS algorithm \citep{nocedal2006numerical}:
\begin{equation}\label{eq:nlls_problem}
\argmin_{\boldsymbol{\beta}} \ \frac{1}{N} \left\| S - \exp \left( \mathbf{X} \boldsymbol{\beta} \right) \right\|_2^2 ,
\end{equation}
where $N$ is the number of acquisitions and $\left\| \ \right\|_2$ denotes L2-norm. Initial positions were computed by solving the problem with ordinary least squares. The result of the fit was considered successful if $\text{AKC}(\mathbf{n}) \geq 0$ for all $\mathbf{n}$ in a 45-point 8-design \citep{hardin1996mclaren, henriques2021diffusional}.

\subsubsection{Prediction of kurtosis maps}\label{step_2}

Using Scikit-learn \citep{pedregosa2011scikit}, a fully-connected feed-forward neural network with rectified linear units as activation functions was trained to predict MK, AK, and RK from data in voxels where the standard NLLS fit was successful. The network had an input layer size equal to the number of acquisitions, three hidden layers with 50 neurons each, and an output layer with size 3. Initial weights were randomly generated with a hard-coded pseudorandom number generator seed. Mean squared error was used as the loss function. Adam \citep{kingma2014adam} was used as the optimizer. Batch size was 200. The training was stopped when the loss did not improve by more than $10^{-4}$ for 10 consecutive epochs. After training, the network was used to predict MK, AK, and RK in all voxels. Training the network was repeated for each scan.

\subsubsection{Regularized non-linear least squares}\label{step_3}
ut
The final parameter estimates were obtained by solving the following optimization problem using the BFGS algorithm:
\begin{equation}
\begin{split}
\argmin_{\boldsymbol{\beta}} \Bigg( \ & \frac{1}{N} \left\| S - \exp \left( \mathbf{X} \boldsymbol{\beta} \right) \right\|_2^2 + \alpha \Big( \left( \hat{\text{MK}} - m(\boldsymbol{\beta})\right) ^ 2 \\ & + \left( \hat{\text{AK}} - a(\boldsymbol{\beta})\right) ^ 2 + \left( \hat{\text{RK}} - r(\boldsymbol{\beta}) \right) ^ 2 \Big) \Bigg) , 
\end{split}
\end{equation}
where $\hat{\text{AK}}$, $\hat{\text{MK}}$, and $\hat{\text{RK}}$ are the predicted kurtosis values, $\alpha$ is a constant controlling the magnitude of the regularization terms, and $m$, $a$, and $r$ are functions for numerically computing mean, axial, and radial kurtosis, respectively \citep{henriques2021diffusional}. Initial positions corresponded to axially symmetric diffusion and kurtosis tensors \citep{hansen2016fast} aligned with the principal diffusion direction and with plausible magnitudes computed from the results of the steps described in sections \ref{step_1} and \ref{step_2}. Following \cite{henriques2021toward}, we used $\alpha = 0.1 \cdot \text{MSE}_\text{NLLS} / \text{MSE}_\text{MK}$, where $\text{MSE}_\text{NLLS}$ is the median squared error of the standard NLLS fit and $\text{MSE}_\text{MK}$ and is the median squared error of the MK prediction.

\subsection{Image acquisition}

Ten healthy volunteers (5 females, 5 males, ages between 26 and 60 years) were scanned on a Siemens Magnetom Prisma 3T (Siemens Healthcare, Erlangen, Germany) with a maximum gradient strength of 80 mT/m, maximum slew rate of 200 T/m/s, and 64-channel head coil at three scan sites: Cardiff University Brain Research Imaging Centre (CUBRIC), Cardiff, United Kingdom; Wolfson Brain Imaging Centre (WBIC), Cambridge, United Kingdom; Great Ormond Street Hospital (GOSH), London, United Kingdom. Ethical approval was given by the research ethics committee and the volunteers gave written and informed consent prior to the scans.

Each scan was performed using the same clinical protocol. Echo-planar multiband PGSE was used with the following parameter values: diffusion time ($\Delta$) = 28.7 ms; diffusion encoding time ($\delta$) = 16.7 ms; b-values = 1,000 and 2,200 s/mm$^2$; 60 diffusion encoding directions distributed uniformly over half a sphere for both b-values; 13 images without diffusion weighting; echo time (TE) = 60 ms; repetition time (TR) = 3,050 ms; FOV = 220 $\times$ 220 ms; voxel size = 2 $\times$ 2 $\times$ 2 mm$^3$; slice gap = 0.2 mm; 66 slices; phase partial Fourier = 6/8. In addition, one image was acquired without diffusion-weighting and the phase encoding direction reversed.

\subsection{Image preprocessing}

Dipy \citep{garyfallidis2014dipy} was used to denoise the raw images using patch2self \citep{fadnavis2020patch}, a self-supervised algorithm that learns to separate signal from noise without model assumptions using oversampled data. MRtrix3 \citep{tournier2019mrtrix3} was used to reduce Gibbs ringing artefacts with a sub-voxel shift algorithm \citep{kellner2016gibbs} and to correct for distortions induced by motion, susceptibility, and eddy currents using FSL's topup and eddy \citep{andersson2016integrated, smith2004advances}.

\subsection{Image segmentation}

TractSeg's pre-trained artificial neural network \citep{wasserthal2018tractseg} was used to segment white matter tracts. The following tracts were included in the analysis: arcuate fascicle (AF); anterior thalamic radiation (ATR); comissure anterior (CA); corpus callosum (CC) and CC divided into rostrum (CC$_1$), genu (CC$_2$), rostral body (CC$_3$), anterior midbody (CC$_4$), posterior midbody (CC$_5$), isthmus (CC$_6$), and splenium (CC$_7$); cingulum (CG); corticospinal tract (CST); fronto-pontine tract (FPT); fornix (FX); inferior cerebellar peduncle (ICP); inferior occipito-frontal facicle (IFO); inferior longitudinal fascicle (ILF); middle cerebellar peduncle (MCP); middle longitudinal fascicle (MLF); optic radiation (OR); parieto-occipital pontine (POPT); superior cerebellar peduncle (SCP); superior longitudinal fascicle (SLF) I, II, and III; superior thalamic radiation (STR); uncinate fascicle (UF). TractSeg's probabilistic outputs were converted into binary masks by using a threshold of 0.9.

\subsection{Statistical analysis}

To quantify the reproducibility of mean values of DTI and DKI parameters in different tracts, a random-effects model was constructed:
\begin{equation}\label{eq:stat_model}
x_{ij} = \mu + \nu_i + e_{ij} ,
\end{equation}
where $x_{ij}$ is the value of microstructural metric $x$ for subject $i$ in scan $j$, $\mu$ is the grand mean of $x$, $\nu_i$ is the subject-level random effect, and $e_{ij}$ is the measurement error. It is assumed that $\nu_i$ and $e_{ij}$ are mutually independent and normally distributed: $\nu_i \sim \mathcal{N}\left(0, \sigma_\nu^2 \right)$, $e_{ij} \sim \mathcal{N}\left(0, \sigma_e^2 \right)$. Point estimates and 95\% confidence intervals were estimated for the model parameter using lme4 \citep{bates2014fitting, team2013r}.  A random-effects model was chosen because a mixed-effects model with a scanner-level fixed effect showed that the scanner-level fixed effect was not significant. Coefficients of variation (CV) quantifying between-subject variability and measurement error were calculated according to the following equations:
\begin{equation}
\text{CV}_\nu = \frac{\sigma_\nu}{\mu} \cdot 100\%
\end{equation}
\begin{equation}
\text{CV}_e = \frac{\sigma_e}{\mu} \cdot 100\%
\end{equation}


\section{Results}

The artificial neural network learned the mapping from data to kurtosis metrics very well. The coefficient of determination was 0.99 for MK, 0.97 for AK, and 0.98 for RK when calculated over all training data.

\begin{figure}[h!]
  \centering
  \includegraphics[width=.9\linewidth]{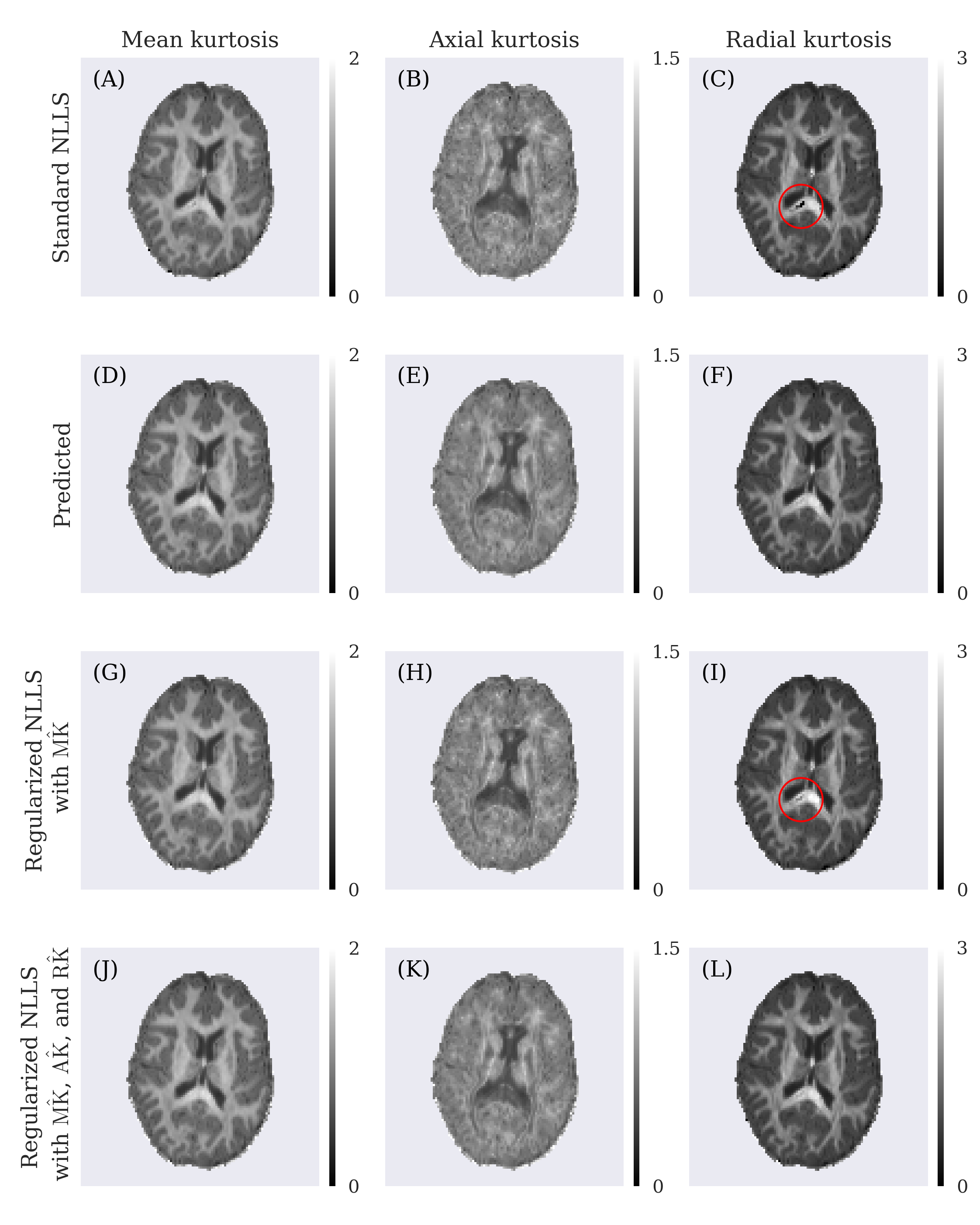}
  \caption{DKI parameter maps computed using standard NLLS (A-C), predicted by the artificial neural networks (D-F), computed using regularized NLLS with $\hat{\text{MK}}$ (G-I), and computed using regularized NLLS with $\hat{\text{MK}}$, $\hat{\text{AK}}$, and $\hat{\text{RK}}$ (J-L). The red circles highlight problematic voxels with implausibly low kurtosis values.}
  \label{fig:maps}
\end{figure}

Figure \ref{fig:maps} shows example DKI parameter maps computed using standard NLLS (A-C), predicted by the artificial neural network (D-F), computed using regularized NLLS with $\hat{\text{MK}}$ (G-I), and computed using regularized NLLS with $\hat{\text{MK}}$, $\hat{\text{AK}}$, and $\hat{\text{RK}}$ (J-L). Standard NLLS produced poor results in several voxels in the splenium of the corpus callosum, which can be seen clearly in the RK map. Regularized NLLS with $\hat{\text{MK}}$ produced clean MK maps but produced some implausible RK values in the splenium of the corpus callosum, motivating the addition of more regularization terms. Indeed, regularized NLLS with $\hat{\text{MK}}$, $\hat{\text{AK}}$, and $\hat{\text{RK}}$ did not suffer from this issue.

\begin{figure}[h!]
  \centering
  \includegraphics[width=.9\linewidth]{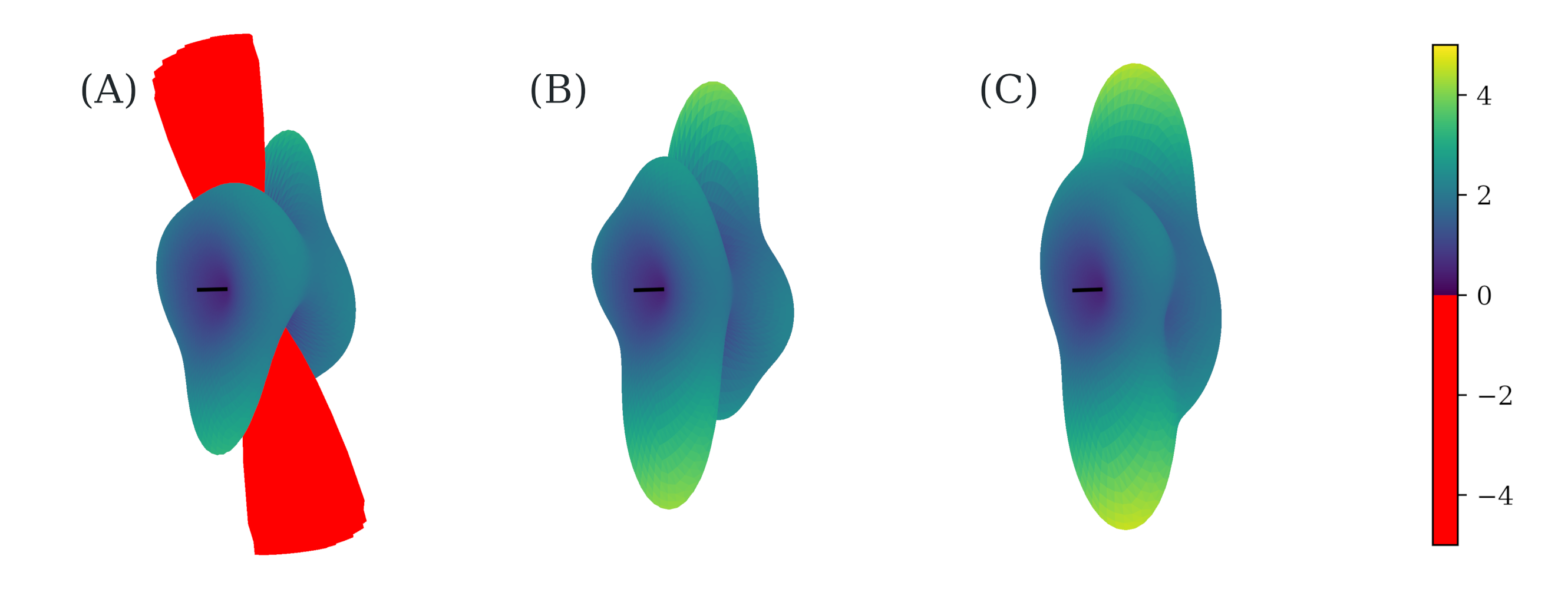}
  \caption{Visualization of a kurtosis tensor in the splenium of the corpus callosum computed using standard NLLS (A), computed using regularized NLLS with $\hat{\text{MK}}$ (B), and computed using regularized NLLS with $\hat{\text{MK}}$, $\hat{\text{AK}}$, and $\hat{\text{RK}}$ (C). AK = 0.52 for all three tensors. MK = 1.3 for tensor A and 1.6 for tensors B and C. RK = -1.3 for tensor A, RK = 1.1 for tensor B, and RK = 2.0 for tensor C. The black line denotes the principal diffusion direction. The shown surfaces were generated by evaluating AKC along different directions. Negative values were clipped at -4.}
  \label{fig:tensor-comparison}
\end{figure}

\begin{figure}
  \centering
  \includegraphics[width=.99\linewidth]{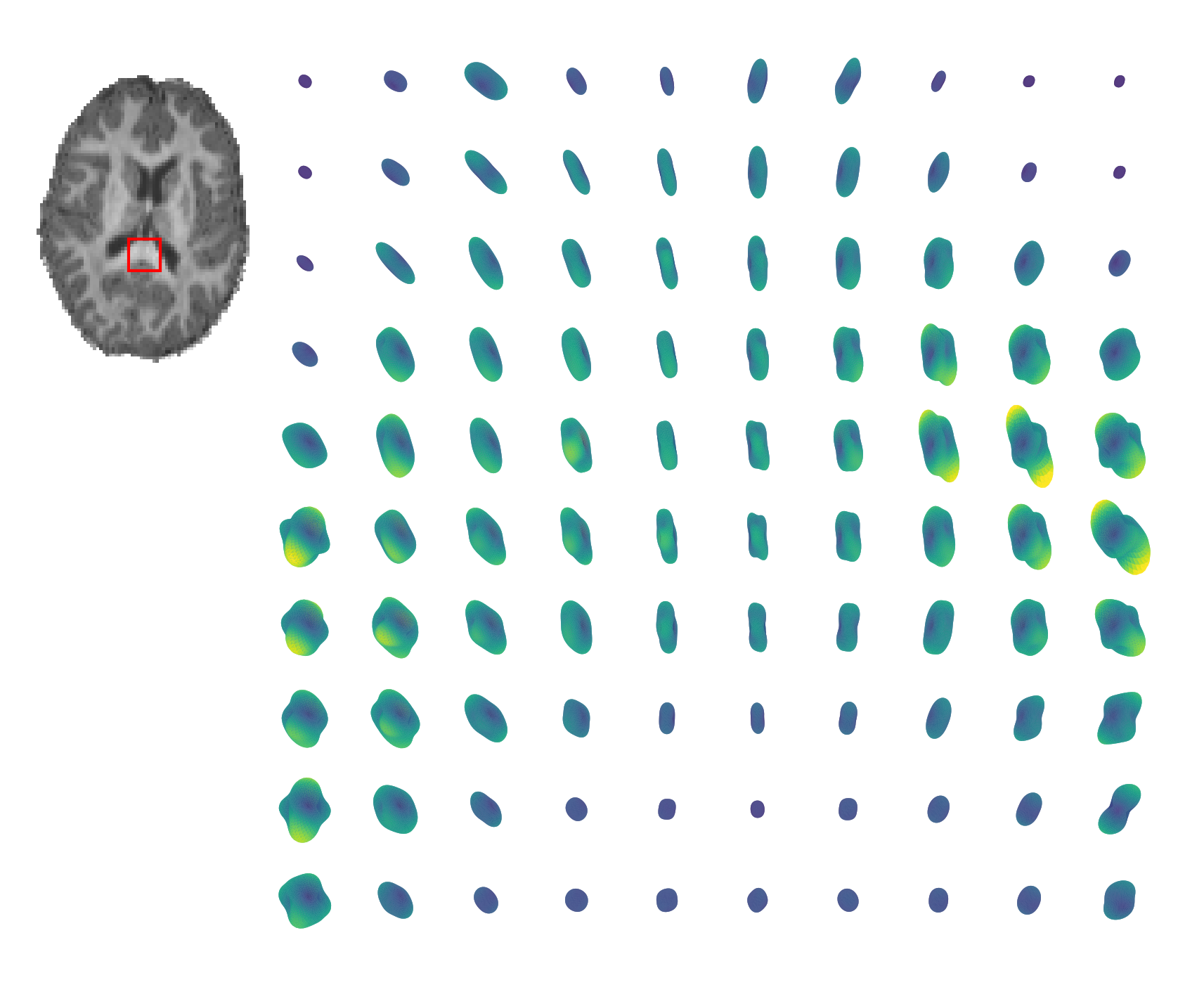}
  \caption{Visualization of the geometry of kurtosis tensors estimated using the proposed algorithm in 100 voxels near the splenium of the corpus callosum where the standard NLLS fit often fails. The shown surfaces were generated by evaluating AKC along different directions.}
  \label{fig:tensors}
\end{figure}

Importantly, our algorithm estimates full tensors and thus provides more information about tissue microstructure than just the predicted parameter maps. Visualizations of the estimated tensors in one of the problematic voxels in the splenium of the corpus callosum are provided in Figure \ref{fig:tensor-comparison} which shows AKC values along different directions. In voxels containing tightly packed aligned axons, diffusion is most restricted along the directions perpendicular to the axons and AKC along these directions should be relatively high. However, due to the complicated geometry of the kurtosis tensor, the estimated tensor can correspond to a correct value of MK and an implausibly low value of RK. A visualization of the geometry of 100 kurtosis tensors near the splenium of the corpus callosum estimated using our algorithm is provided in Figure \ref{fig:tensors}.

\begin{figure}
  \centering
  \includegraphics[width=.99\linewidth]{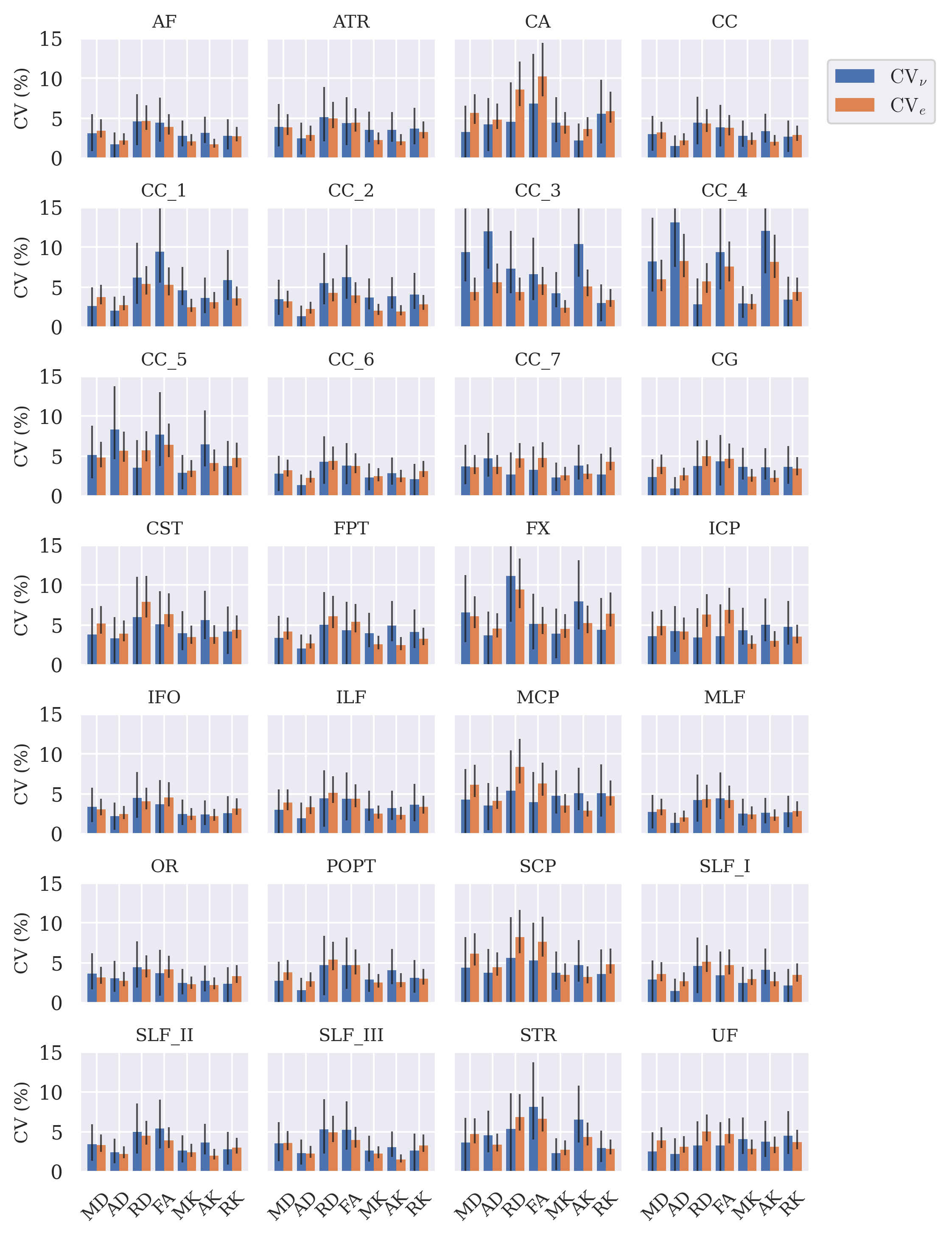}
  \caption{Estimated CV$_\nu$ and CV$_e$ for DTI and DKI parameters in different tracts. The black bars represent 95\% confidence intervals.}
  \label{fig:CV}
\end{figure}

The results of the reproducibility analysis are shown in Figure \ref{fig:CV}. Overall, the reproducibility of DKI metrics, as measured by the estimated variability due to measurement error ($\text{CV}_e$), was comparable to DTI metrics. The relative differences between the metrics' reproducibility depended on the tract. The ratio betwen between-subject variability and measurement error also varied greatly between the tracts. In large tracts, such as the corpus callosum and longitudinal fascicles, CV values for all metrics were below 5\%. In some smaller tracts, such as the fornix and subparts of the corpus callosum, CV values exceeded 5\% for some diffusivity and kurtosis metrics.

All the computational steps including training required to compute whole-brain DTI and DKI parameter maps took less than 10 minutes per scan on an Intel Xeon CPU E5-1620 v3 3.50GHz x 8.

\section{Discussion}

The aim of this study was to investigate the reproducibility of DKI parameters in human white matter using regularized NLLS. Based on MK's independence of RK and AK, we hypothesized that imposing regularization only on MK as proposed by \cite{henriques2021toward} may not be enough for robust estimation of the kurtosis tensor. We discovered that, despite providing signifcantly better results than standard NLLS, regularizing only MK can result in kurtosis tensor estimates that have implausibly low RK. This is due to the fact that the kurtosis tensor can have a complicated shape that corresponds to the correct value of MK but an implausibly low value of RK. This issue, highlighted in figures \ref{fig:maps} and \ref{fig:tensor-comparison}, motivated us to also regularize kurtosis tensor magnitude along and perpendicular to the principal diffusion direction. Our algorithm has been made publicly available in an easy-to-use open-source Python package to promote reproducible science and to encourage other researchers to further improve the method.

We chose to use feed-forward artificial neural networks to learn the mapping from data to DKI parameter maps because they are universal function approximators \citep{hornik1989multilayer} that do not require manual feature engineering. Depending on the resolution, whole-brain dMRI scans typically contain between tens of thousands and over a million voxels, providing sufficient training data. Our algorithm assumes that the training data, i.e., the voxels where the standard NLLS fit did not produce negative AKC values, contains a wide range of different tissue types with varying microstructural properties, making the trained neural networks generalizable. This is justified since we know that the standard NLLS fit tends to fail in specific edge cases that make up less than 1\% of the whole brain, given high enough data quality. Although the artificial neural networks learned the mapping from data to kurtosis metrics very well on our data, worse predictions are expected if the maps are noisier due to lower data quality. This can be adjusted by changing the network architecture or hypermarameter values.

The results of our statistical analysis show that the reproducibility of the DKI parameters estimated using our method is comparable to DTI parameters when applied on data acquired using a modern clinical scan protocol. However, it must be noted that our data was of high quality and preprocessed using state-of-the-art preprocessing methods. It is expected that our algorithm is less robust if applied on low-quality data.

A limitation of this work is that we applied our algorithm only on imaging data for which the ground truth is unknown. Therefore, possible biases must be investigated in future studies. Nevertheless, the high prediction accuracy suggests that we are not introducing significant biases compared to the standard NLLS fit. Also, since our method is based on the BFGS algorithm, it can end up in a local minimum and be sensitive to the initial position. Finally, the value of the parameter $\alpha$, a constant controlling the relative weights of data points and predicted values in the objective function, must be manually chosen and its optimal value may depend on the acquisition protocol.

\newpage

\section*{Figures}

\FloatBarrier

\newpage

\newpage

\newpage

\FloatBarrier

\newpage

\section*{Data and code availability}

The code is publicly available at \url{https://github.com/kerkelae/dkmri}.

\section*{Author credit}

\textbf{Leevi Kerkelä}: Conceptualization, Data curation, Formal analysis, Investigation, Methodology, Project administration, Software, Validation, Visualization, Writing - Original Draft. \textbf{Kiran Seunarine}: Data Curation, Investigation, Writing - Review \& Editing. \textbf{Rafael Neto Henriques}: Methodology, Validation, Writing - Review \& Editing. \textbf{Jonathan D. Clayden}: Methodology, Writing - Review \& Editing. \textbf{Chris A. Clark}: Funding acquisition, Project administration, Resources, Writing - Review \& Editing.

\section*{Acknowledgements}

This study has received funding from the European Union’s Horizon 2020 research and innovation programme under grant agreement No 847826. The authors would like to thank John Evans, Marta Correia, and Steve Eldridge for assisting with data acquisition.

\newpage

\bibliography{main}

\begin{thebibliography}{}

\bibitem[Andersson and Sotiropoulos, 2016]{andersson2016integrated}
Andersson, J.~L. and Sotiropoulos, S.~N. (2016).
\newblock An integrated approach to correction for off-resonance effects and
  subject movement in diffusion mr imaging.
\newblock {\em Neuroimage}, 125:1063--1078.

\bibitem[Barnea-Goraly et~al., 2005]{barnea2005white}
Barnea-Goraly, N., Menon, V., Eckert, M., Tamm, L., Bammer, R., Karchemskiy,
  A., Dant, C.~C., and Reiss, A.~L. (2005).
\newblock White matter development during childhood and adolescence: a
  cross-sectional diffusion tensor imaging study.
\newblock {\em Cerebral cortex}, 15(12):1848--1854.

\bibitem[Basser et~al., 1994]{basser1994mr}
Basser, P.~J., Mattiello, J., and LeBihan, D. (1994).
\newblock Mr diffusion tensor spectroscopy and imaging.
\newblock {\em Biophysical journal}, 66(1):259--267.

\bibitem[Bates et~al., 2014]{bates2014fitting}
Bates, D., M{\"a}chler, M., Bolker, B., and Walker, S. (2014).
\newblock Fitting linear mixed-effects models using lme4.
\newblock {\em arXiv preprint arXiv:1406.5823}.

\bibitem[Blumenfeld-Katzir et~al., 2011]{blumenfeld2011diffusion}
Blumenfeld-Katzir, T., Pasternak, O., Dagan, M., and Assaf, Y. (2011).
\newblock Diffusion mri of structural brain plasticity induced by a learning
  and memory task.
\newblock {\em PloS one}, 6(6):e20678.

\bibitem[Bradbury et~al., 2018]{jax2018github}
Bradbury, J., Frostig, R., Hawkins, P., Johnson, M.~J., Leary, C., Maclaurin,
  D., Necula, G., Paszke, A., Vander{P}las, J., Wanderman-{M}ilne, S., and
  Zhang, Q. (2018).
\newblock {JAX}: composable transformations of {P}ython+{N}um{P}y programs.

\bibitem[Fadnavis et~al., 2020]{fadnavis2020patch}
Fadnavis, S., Batson, J., and Garyfallidis, E. (2020).
\newblock Patch2self: Denoising diffusion mri with self-supervised learning​.
\newblock In Larochelle, H., Ranzato, M., Hadsell, R., Balcan, M.~F., and Lin,
  H., editors, {\em Advances in Neural Information Processing Systems},
  volume~33, pages 16293--16303. Curran Associates, Inc.

\bibitem[Falangola et~al., 2008]{falangola2008age}
Falangola, M.~F., Jensen, J.~H., Babb, J.~S., Hu, C., Castellanos, F.~X.,
  Di~Martino, A., Ferris, S.~H., and Helpern, J.~A. (2008).
\newblock Age-related non-gaussian diffusion patterns in the prefrontal brain.
\newblock {\em Journal of Magnetic Resonance Imaging: An Official Journal of
  the International Society for Magnetic Resonance in Medicine},
  28(6):1345--1350.

\bibitem[Garyfallidis et~al., 2014]{garyfallidis2014dipy}
Garyfallidis, E., Brett, M., Amirbekian, B., Rokem, A., Van Der~Walt, S.,
  Descoteaux, M., and Nimmo-Smith, I. (2014).
\newblock Dipy, a library for the analysis of diffusion mri data.
\newblock {\em Frontiers in neuroinformatics}, 8:8.

\bibitem[Grieve et~al., 2007]{grieve2007cognitive}
Grieve, S.~M., Williams, L.~M., Paul, R.~H., Clark, C.~R., and Gordon, E.
  (2007).
\newblock Cognitive aging, executive function, and fractional anisotropy: a
  diffusion tensor mr imaging study.
\newblock {\em American Journal of Neuroradiology}, 28(2):226--235.

\bibitem[Hansen et~al., 2016]{hansen2016fast}
Hansen, B., Shemesh, N., and Jespersen, S.~N. (2016).
\newblock Fast imaging of mean, axial and radial diffusion kurtosis.
\newblock {\em Neuroimage}, 142:381--393.

\bibitem[Hardin and Sloane, 1996]{hardin1996mclaren}
Hardin, R.~H. and Sloane, N.~J. (1996).
\newblock Mclaren’s improved snub cube and other new spherical designs in
  three dimensions.
\newblock {\em Discrete \& Computational Geometry}, 15(4):429--441.

\bibitem[Henriques et~al., 2021a]{henriques2021diffusional}
Henriques, R.~N., Correia, M.~M., Marrale, M., Huber, E., Kruper, J., Koudoro,
  S., Yeatman, J.~D., Garyfallidis, E., and Rokem, A. (2021a).
\newblock Diffusional kurtosis imaging in the diffusion imaging in python
  project.
\newblock {\em Frontiers in Human Neuroscience}, page 390.

\bibitem[Henriques et~al., 2021b]{henriques2021toward}
Henriques, R.~N., Jespersen, S.~N., Jones, D.~K., and Veraart, J. (2021b).
\newblock Toward more robust and reproducible diffusion kurtosis imaging.
\newblock {\em Magnetic Resonance in Medicine}.

\bibitem[Hornik et~al., 1989]{hornik1989multilayer}
Hornik, K., Stinchcombe, M., and White, H. (1989).
\newblock Multilayer feedforward networks are universal approximators.
\newblock {\em Neural networks}, 2(5):359--366.

\bibitem[Hui et~al., 2012]{hui2012stroke}
Hui, E.~S., Fieremans, E., Jensen, J.~H., Tabesh, A., Feng, W., Bonilha, L.,
  Spampinato, M.~V., Adams, R., and Helpern, J.~A. (2012).
\newblock Stroke assessment with diffusional kurtosis imaging.
\newblock {\em Stroke}, 43(11):2968--2973.

\bibitem[Jensen and Helpern, 2010]{jensen2010mri}
Jensen, J.~H. and Helpern, J.~A. (2010).
\newblock Mri quantification of non-gaussian water diffusion by kurtosis
  analysis.
\newblock {\em NMR in Biomedicine}, 23(7):698--710.

\bibitem[Jensen et~al., 2005]{jensen2005diffusional}
Jensen, J.~H., Helpern, J.~A., Ramani, A., Lu, H., and Kaczynski, K. (2005).
\newblock Diffusional kurtosis imaging: the quantification of non-gaussian
  water diffusion by means of magnetic resonance imaging.
\newblock {\em Magnetic Resonance in Medicine: An Official Journal of the
  International Society for Magnetic Resonance in Medicine}, 53(6):1432--1440.

\bibitem[Johansen-Berg and Behrens, 2013]{johansen2013diffusion}
Johansen-Berg, H. and Behrens, T.~E. (2013).
\newblock {\em Diffusion MRI: from quantitative measurement to in vivo
  neuroanatomy}.
\newblock Academic Press.

\bibitem[Kellner et~al., 2016]{kellner2016gibbs}
Kellner, E., Dhital, B., Kiselev, V.~G., and Reisert, M. (2016).
\newblock Gibbs-ringing artifact removal based on local subvoxel-shifts.
\newblock {\em Magnetic resonance in medicine}, 76(5):1574--1581.

\bibitem[Kingma and Ba, 2014]{kingma2014adam}
Kingma, D.~P. and Ba, J. (2014).
\newblock Adam: A method for stochastic optimization.
\newblock {\em arXiv preprint arXiv:1412.6980}.

\bibitem[Kiselev, 2010]{kiselev2010cumulant}
Kiselev, V.~G. (2010).
\newblock The cumulant expansion: an overarching mathematical framework for
  understanding diffusion nmr.
\newblock {\em Diffusion MRI}, pages 152--168.

\bibitem[Kuder et~al., 2012]{kuder2012advanced}
Kuder, T.~A., Stieltjes, B., Bachert, P., Semmler, W., and Laun, F.~B. (2012).
\newblock Advanced fit of the diffusion kurtosis tensor by directional
  weighting and regularization.
\newblock {\em Magnetic resonance in medicine}, 67(5):1401--1411.

\bibitem[Lam et~al., 2015]{lam2015numba}
Lam, S.~K., Pitrou, A., and Seibert, S. (2015).
\newblock Numba: A llvm-based python jit compiler.
\newblock In {\em Proceedings of the Second Workshop on the LLVM Compiler
  Infrastructure in HPC}, pages 1--6.

\bibitem[Neto~Henriques, 2018]{neto2018advanced}
Neto~Henriques, R. (2018).
\newblock {\em Advanced Methods for Diffusion MRI Data Analysis and their
  Application to the Healthy Ageing Brain}.
\newblock PhD thesis, University of Cambridge.

\bibitem[Nocedal and Wright, 2006]{nocedal2006numerical}
Nocedal, J. and Wright, S. (2006).
\newblock {\em Numerical optimization}.
\newblock Springer Science \& Business Media.

\bibitem[Pedregosa et~al., 2011]{pedregosa2011scikit}
Pedregosa, F., Varoquaux, G., Gramfort, A., Michel, V., Thirion, B., Grisel,
  O., Blondel, M., Prettenhofer, P., Weiss, R., Dubourg, V., et~al. (2011).
\newblock Scikit-learn: Machine learning in python.
\newblock {\em the Journal of machine Learning research}, 12:2825--2830.

\bibitem[Price et~al., 2017]{price2017age}
Price, D., Tyler, L.~K., Henriques, R.~N., Campbell, K.~L., Williams, N.,
  Treder, M.~S., Taylor, J.~R., and Henson, R.~N. (2017).
\newblock Age-related delay in visual and auditory evoked responses is mediated
  by white-and grey-matter differences.
\newblock {\em Nature communications}, 8(1):1--12.

\bibitem[{R Core Team}, 2013]{team2013r}
{R Core Team} (2013).
\newblock R: A language and environment for statistical computing.

\bibitem[Shafto et~al., 2014]{shafto2014cambridge}
Shafto, M.~A., Tyler, L.~K., Dixon, M., Taylor, J.~R., Rowe, J.~B., Cusack, R.,
  Calder, A.~J., Marslen-Wilson, W.~D., Duncan, J., Dalgleish, T., et~al.
  (2014).
\newblock The cambridge centre for ageing and neuroscience (cam-can) study
  protocol: a cross-sectional, lifespan, multidisciplinary examination of
  healthy cognitive ageing.
\newblock {\em BMC neurology}, 14(1):1--25.

\bibitem[Smith et~al., 2004]{smith2004advances}
Smith, S.~M., Jenkinson, M., Woolrich, M.~W., Beckmann, C.~F., Behrens, T.~E.,
  Johansen-Berg, H., Bannister, P.~R., De~Luca, M., Drobnjak, I., Flitney,
  D.~E., et~al. (2004).
\newblock Advances in functional and structural mr image analysis and
  implementation as fsl.
\newblock {\em Neuroimage}, 23:S208--S219.

\bibitem[Tabesh et~al., 2011]{tabesh2011estimation}
Tabesh, A., Jensen, J.~H., Ardekani, B.~A., and Helpern, J.~A. (2011).
\newblock Estimation of tensors and tensor-derived measures in diffusional
  kurtosis imaging.
\newblock {\em Magnetic resonance in medicine}, 65(3):823--836.

\bibitem[Tournier et~al., 2019]{tournier2019mrtrix3}
Tournier, J.-D., Smith, R., Raffelt, D., Tabbara, R., Dhollander, T., Pietsch,
  M., Christiaens, D., Jeurissen, B., Yeh, C.-H., and Connelly, A. (2019).
\newblock Mrtrix3: A fast, flexible and open software framework for medical
  image processing and visualisation.
\newblock {\em Neuroimage}, 202:116137.

\bibitem[Van~Cauter et~al., 2012]{van2012gliomas}
Van~Cauter, S., Veraart, J., Sijbers, J., Peeters, R.~R., Himmelreich, U.,
  De~Keyzer, F., Van~Gool, S.~W., Van~Calenbergh, F., De~Vleeschouwer, S.,
  Van~Hecke, W., et~al. (2012).
\newblock Gliomas: diffusion kurtosis mr imaging in grading.
\newblock {\em Radiology}, 263(2):492--501.

\bibitem[Van~Essen et~al., 2013]{van2013wu}
Van~Essen, D.~C., Smith, S.~M., Barch, D.~M., Behrens, T.~E., Yacoub, E.,
  Ugurbil, K., Consortium, W.-M.~H., et~al. (2013).
\newblock The wu-minn human connectome project: an overview.
\newblock {\em Neuroimage}, 80:62--79.

\bibitem[Wasserthal et~al., 2018]{wasserthal2018tractseg}
Wasserthal, J., Neher, P., and Maier-Hein, K.~H. (2018).
\newblock Tractseg-fast and accurate white matter tract segmentation.
\newblock {\em NeuroImage}, 183:239--253.

\bibitem[Zhang et~al., 2009]{zhang2009white}
Zhang, Y., Schuff, N., Du, A.-T., Rosen, H.~J., Kramer, J.~H., Gorno-Tempini,
  M.~L., Miller, B.~L., and Weiner, M.~W. (2009).
\newblock White matter damage in frontotemporal dementia and alzheimer's
  disease measured by diffusion mri.
\newblock {\em Brain}, 132(9):2579--2592.

\bibitem[Zhuo et~al., 2012]{zhuo2012diffusion}
Zhuo, J., Xu, S., Proctor, J.~L., Mullins, R.~J., Simon, J.~Z., Fiskum, G., and
  Gullapalli, R.~P. (2012).
\newblock Diffusion kurtosis as an in vivo imaging marker for reactive
  astrogliosis in traumatic brain injury.
\newblock {\em Neuroimage}, 59(1):467--477.

\end{thebibliography}
\bibliographystyle{apalike}

\end{document}